# Nonlinearity Attack against the Kirchhoff-Law-Johnson-Noise (KLJN) Secure Key Exchange Protocol

Christiana Chamon [1*] and Laszlo B. Kish [2]

Department of Electrical and Computer Engineering, Dwight Look College of Engineering, Texas A&M University, College Station, TX 77843-3128, USA; laszlokish@tamu.edu
* Correspondence: cschamon@tamu.edu

**Abstract:** This paper introduces a new attack against the Kirchhoff-Law-Johnson-Noise (KLJN) secure key exchange scheme. The attack is based on the nonlinearity of the noise generators. We explore the effect of total distortion ($TD$) at the second order ($D_2$), third order ($D_3$), and a combination of the second and third orders ($D_{2,3}$) on the security of the KLJN scheme. It is demonstrated that a $TD$ as little as 1% results in a notable power flow along the information channel, which leads to a significant information leak. We also show that decreasing the effective temperature (that is, the wire voltage) and, in this way reducing nonlinearity, results in the KLJN scheme approaching perfect security.

## 1. Introduction

*1.1. Secure Communications*

Secure communications involve communicating parties Alice and Bob exchanging messages over a public channel. In the symmetric-key protocol, they use the same secure key and ciphers to perform encryption and decryption [1]. Thus there is a demand for a secure key exchange, or the generation and distribution of the secure key over the information channel. This is usually the most demanding process because the secure key exchange is a secure communication itself.

Note, the secure key exchange must invoke Kerckhoffs's Principle/Shannon's Maxim [2]: Eve knows everything there is to know about the key exchange system, except for the key.

*1.2. Conditional Security*

A secure key exchange system can be either conditionally or unconditionally secure. The algorithmic secure key exchange systems of today rely on limited computational power and mathematically hard problems to solve. These are regarded as conditionally secure because Eve has all the data needed to crack the key and only the available computational power provides the security over a limited time interval. These systems are not future-proof as technology and algorithms can evolve in unexpected ways . This raises the demand for unconditionally (information theoretically) secure key exchange systems where, in the situation of perfect security, Eve has not useful data for cracking the secure key and computational power is irrelevant.

*1.3. Unonditional Security*

The security of known unconditionally secure key exchange systems is guaranteed by the laws of physics. The distinction that these systems provide is that no matter how advanced Eve's equipment is, her information entropy does not decrease, even after an arbitrary attack or an unlimited amount of time. There are currently two types of unconditionally secure key exchange systems: quantum key distribution (QKD) [3-38] and the Kirchhoff-law-Johnson-noise (KLJN) scheme [39-96]. Important criticisms have been made about QKD in relation to its fundamental claims and the realization in practice [3-38].

Our focus topic, the KLJN scheme, which is the classical physics competitor of QKD, is a statistical physical secure key exchange scheme whose unconditional security is based on the 2nd law of thermodynamics (i.e. the impossibility to build a perpetual motion machine of the second kind) [39-43].

*1.4. The KLJN Secure Key Exchange System*

Figure 1 shows the core of the KLJN scheme. Alice and Bob are connected via a wire (with voltage and current $U_w(t)$ and $I_w(t)$, respectively), which serves as their information channel. They each have identical pairs of resistors $R_H$ and $R_L$ $(R_H > R_L)$ with respective thermal noise voltages $U_{H,A}(t)$ and $U_{L,A}(t)$, and $U_{H,B}(t)$, $U_{L,B}(t)$.

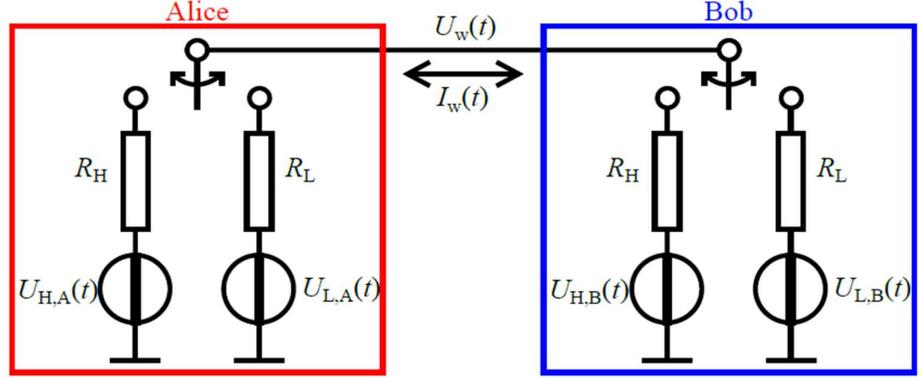

**Figure 1.** The core of the KLJN scheme. Communicating parties Alice and Bob are connected via a wire (with voltage and current $U_w(t)$ and $I_w(t)$, respectively), which serves as their information channel. They each have identical pairs of resistors $R_H$ and $R_L$ $(R_H > R_L)$ with respective thermal noise voltages $U_{H,A}(t)$, $U_{L,A}(t)$, and $U_{H,B}(t)$, $U_{L,B}(t)$.

At the beginning of the bit exchange period, Alice and Bob randomly select one of their resistors to connect to the wire. In the voltage-based protocol, they measure the mean-square voltage of the wire. Theoretically, the mean-square voltage of the wire is given by the Johnson formula,

$$U_w^2 = 4kT_{eff}R_p\Delta f_B, \qquad (1)$$

where $k$ is the Boltzmann constant (1.38 x 10$^{-23}$ J/K), $T_{eff}$ is the publicly-agreed effective temperature (usually $T_{eff} > 10^{12}$ K), $R_p$ is the parallel resultant of Alice's and Bob's chosen resistors $R_A$ and $R_B$, respectively, given by

$$R_p = \frac{R_A R_B}{R_A + R_B}, \qquad (2)$$

and $\Delta f_B$ is the noise bandwidth of the generators emulating the thermal noise.

There are four possible bit situations that can be formed by Alice's and Bob's choices of resistors: HH, LL, LH, and HL. From the Johnson Formula (see Equations 1 and 2), this results in three possible mean-square voltage levels, as illustrated in Figure 2. The HH and LL bit situations are insecure because they render in a distinct mean-square voltage. Alice and Bob discard these periods. The LH and HL bit situations, on the other hand, are secure because they render the same mean-square voltage. Eve cannot differentiate between the LH and HL bit situations, but Alice and Bob can because they know which resistor they have chosen.

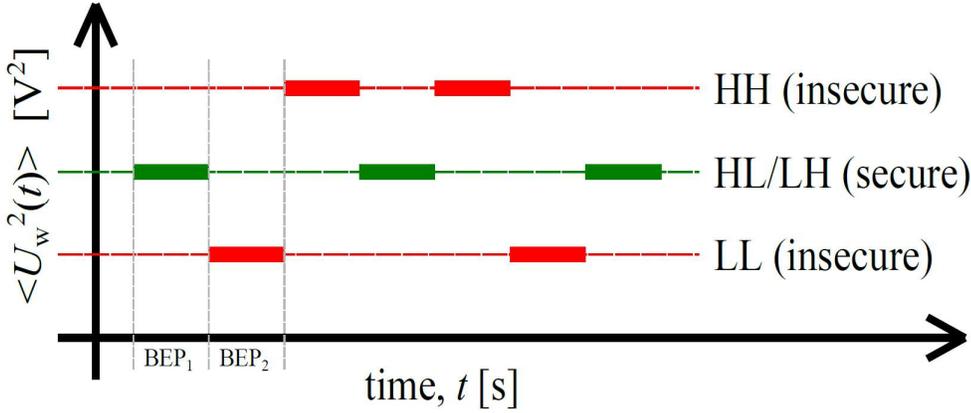

**Figure 2.** The three mean-square voltage levels of the KLJN scheme. The HH and LL bit situations are insecure because they render in a distinct mean-square voltage. Alice and Bob discard these bits.

Several attacks have been proposed against the KLJN scheme [73-96], but each known attack was either conceptually/experimentally incorrect or met with a nullifying defense scheme. In this paper, we propose an attack based on the nonlinear properties of Alice's and Bob's noise generators.

*1.5. Nonlinearity*

The noise generators of Alice and Bob have analog amplifiers as drivers. These have nonlinear characteristics [97]. We can model their output voltage by taking the Taylor Series approximation

$$U^*(t) = A\left[U(t) + BU^2(t) + CU^3(t) + ...\right], \quad (3)$$

where $U^*(t)$ is the output voltage of the generator, $A$ is the linear amplification, $U(t)$ is the input noise voltage, and $B$ and $C$ are the second and third order nonlinearity coefficients, respectively. Nonlinearity obviously distorts the amplitude distribution function and the Gaussianity of the noise sources. Vadai, Mingesz, and Gingl mathematically proved [17] that the KLJN scheme is secure only if the distribution of the noise voltages is Gaussian. Thus nonlinearity is expected to cause information leak in these systems. It is an open question how much is this leak at practical conditions.

In this paper, we explore the effect of nonlinearity at the second order, third order, and a combination of the two orders. We also show that, as we decrease $T_{eff}$, the KLJN scheme approaches perfect security because the nonlinear components get negligibly small due to the reduced noise voltage.

**2. The Nonlinearity Attack**

For illustrative purposes, we use only the second and third order nonlinearities to account for the effects of the even and odd order nonlinearities. To quantify the nonlinearity, we use the total distortion, given by the sum of the normalized mean-square components:

$$\text{TD} = \frac{\sqrt{\left\langle\left[BU^2(t)\right]^2\right\rangle + \left\langle\left[CU^3(t)\right]^2\right\rangle}}{\left\langle\left[U(t)\right]^2\right\rangle}. \quad (4)$$

Eve measures the channel voltage and current, $U_w(t)$ and $I_w(t)$ (see Figure 1) and calculates the net power flow from Alice to Bob,

$$\left\langle P_w(t)\right\rangle = \left\langle I_w(t)U_w(t)\right\rangle, \quad (5)$$

where interpretation of voltage and current polarities are properly chosen for the direction of the power flow. Suppose the following protocol is publicly shared between Alice and Bob:

(i) If the net power flow is greater than zero, Eve surmises that HL is the secure bit situation;

(ii) If the net power flow is less than zero, Eve surmises that LH is the secure bit situation.

For example, in accordance with Equation 1 and 3 we conclude: In the case of positive nonlinear coefficients in Equation 3, the HL case means higher mean-square voltage and higher temperature at Alice's end, thus a positive power flow from Alice to Bob. If Eve extracts a key, she can test that key or its inverse. One of them will be the true key. (For example, with proper negative coefficients, HL can imply a negative power flow, which would lead to the inverse key. If Eve, in accordance with Kerckhoffs's principle, knows the nonlinear coefficient in Equation 3, the inverse operation with the key is not needed.)

**3. Demonstration**

Computer simulations with Matlab measure the information leak with practical nonlinearity parameters in Equation 3. The tests show a significant amount of information leak even with small nonlinearity.

The protocol is as follows:

- For each bit exchange, Eve measures and evaluates the average power at the information channel $\langle P_w(t) \rangle$ (see Equation 5).
- If the result is greater than zero, she guesses that HL is the secure bit situation;
- If the result is less than zero, she guesses that LH is the secure bit situation (see Section 2).
- The process above is independently repeated 1,000 times to obtain the statistics shown.

Out of the linear (Ideal) case, the investigated nonlinear situations are:

*(a)* Case $D_2$ with second-order nonlinearity;
*(b)* Case $D_3$ with third-order nonlinearity;
*(c)* Case $D_{2,3}$ with the combination of the $D_2$ and the $D_3$ cases.

Figure 3 illustrates the IV scatterplots between the wire voltage and current for the Ideal (a), $D_2$ (b), $D_3$ (c), and $D_{2,3}$ (d) situations. The chosen parameters are $R_H$ = 100 kΩ, $R_L$ = 10 kΩ, $T_{eff}$ = $10^{18}$ K, and $\Delta f_B$ = 500 Hz. At $D_2$, $B$ = 6 x $10^{-3}$ and $C$ = 0. At $D_3$, $B$ = 0 and $C$ = 5 x $10^{-5}$. At $D_{2,3}$, $B$ = 1 x $10^{-6}$ and $C$ = 5 x $10^{-5}$. The blue circles represent the HL case, whereas the orange crosses represent the LH case.

The HL and LH situations are statistically indistinguishable in the Ideal (linear) situation, indicating perfect security.

In the $D_2$ case, the HL arrangement has an upward dominance, while the LH has a downward tendency. In the $D_3$ and $D_{2,3}$ cases, the HL situation has a right-diagonal footprint, while the LH situation has a left-diagonal footprint. In conclusion, the nonlinear IU scatterplots indicate lack of security at the given conditions.

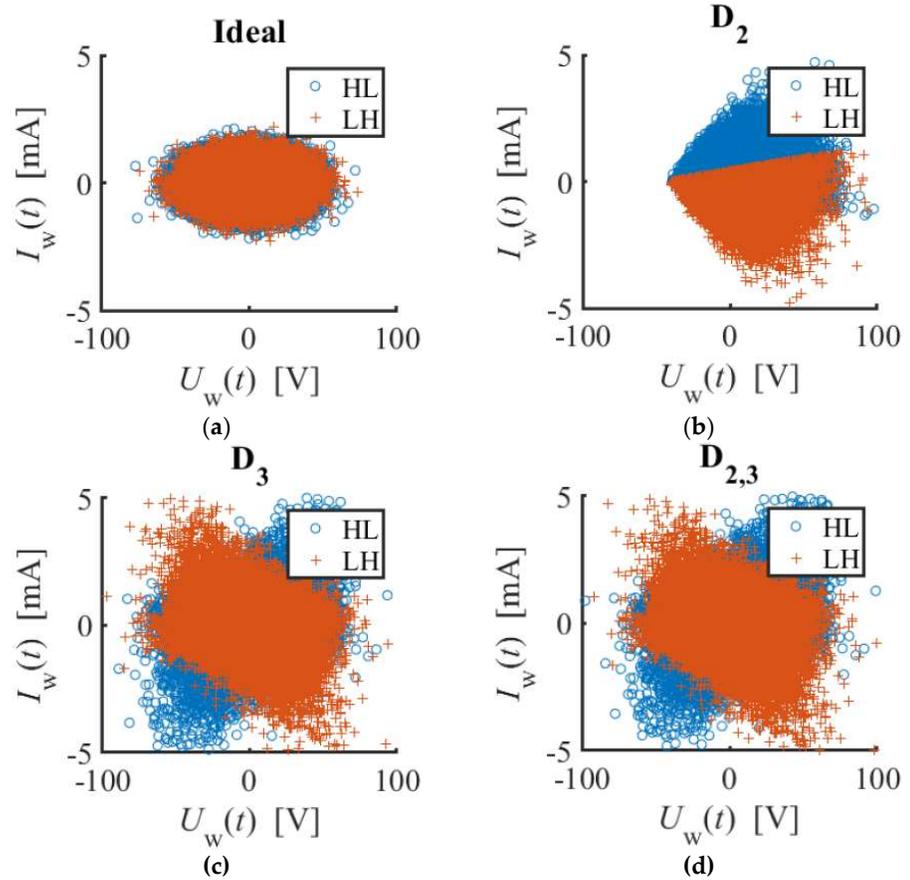

**Figure 3.** The IU scatterplots between the wire voltage and current for the ideal (a), $D_2$ (b), $D_3$ (c), and $D_{2,3}$ (d) situations. The parameters chosen are $R_H$ = 100 kΩ, $R_L$ = 10 kΩ, $T_{eff}$ = $10^{18}$ K, and $\Delta f_B$ = 500 Hz. At $D_2$, B = 6 x $10^{-3}$ and C = 0. At $D_3$, B = 0 and C = 5 x $10^{-5}$. At $D_{2,3}$, B = 1 x $10^{-6}$ and C = 5 x $10^{-5}$. The blue circles represent the HL case, whereas the orange crosses represent the LH case. The HL and LH situations are statistically indistinguishable in the ideal situation. In the $D_2$ case, the HL arrangement has an upward dominance, while the LH has a downward tendency. In the $D_3$ and $D_{2,3}$ cases, the HL situation has a right-diagonal trajectory, while the LH has a left-diagonal trajectory.

Table 1 shows the statistical run for Eve's probability $p$ of correctly guessing the bit situations, and its standard deviation $\sigma$, for four different sample sizes (time steps) $\gamma$. For each nonlinearity situation, the $p$ value increases as $\gamma$ increases, as expected, due to the increasing accuracy of Eve's statistics.

**Table 1.** The statistical sun for Eve's correct-guessing probability $p$ and its standard deviation $\sigma$ for four different sample sizes $\gamma$. For each nonlinearity situation, the $p$ value increases as $\gamma$ increases.

| D | $\gamma$ | $p$ | $\sigma$ |
|---|---|---|---|
| 2 | 10 | 0.5502 | 0.0135 |
|  | 20 | 0.6172 | 0.0203 |
|  | 100 | 0.7498 | 0.0149 |
|  | 1000 | 0.9869 | 0.0042 |
| 3 | 10 | 0.5632 | 0.0159 |
|  | 20 | 0.5982 | 0.0140 |
|  | 100 | 0.7383 | 0.0126 |
|  | 1000 | 0.9831 | 0.0047 |
| 2,3 | 10 | 0.5761 | 0.0114 |
|  | 20 | 0.6106 | 0.0166 |
|  | 100 | 0.7434 | 0.0137 |
|  | 1000 | 0.9855 | 0.0037 |

Varying the effective temperature $T_{\text{eff}}$ resulted in varying the effective voltage on the wire $U_w$ (see Equation 1). The statistical protocol with results shown in Table 1 was repeated for various effective temperatures.

Figure 4 illustrates Eve's correct-guessing probability $p$ (top) and Eve's bit error $\varepsilon$ (bottom), given by

$$\varepsilon = 1 - p, \tag{6}$$

with respect to the effective wire voltage $U_w$ for $D_2$ (a), $D_3$ (b), and $D_{2,3}$ (c) for all sample sizes $\gamma$. As $\gamma$ and $U_w$ (that is the effective temperature) decrease, $p$ approaches perfect security.

Figure 5 shows $p$ and $\varepsilon$ vs. $U_w$ at $\gamma = 1{,}000$ for all the distortions. With the given parameters, convergence toward perfect security happens at $D_2$ before $D_3$ and $D_{2,3}$.

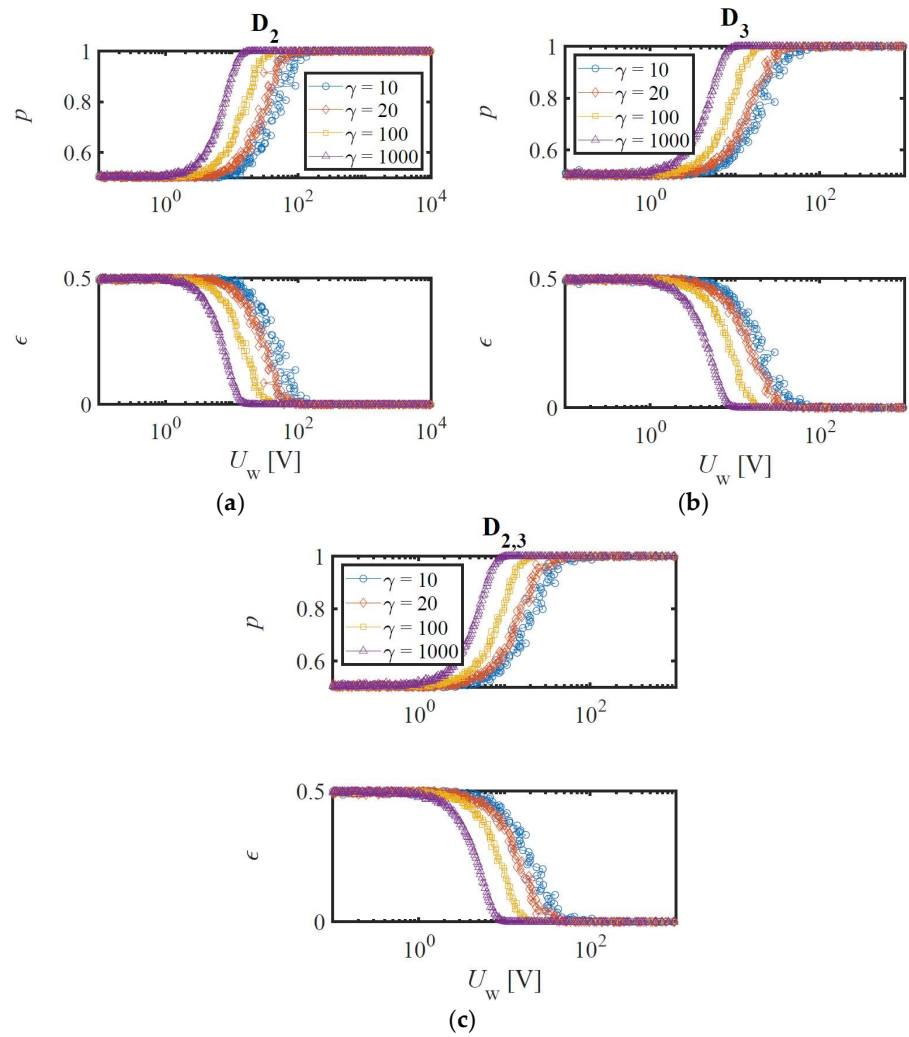

**Figure 4.** Eve's correct-bit-guessing probability $p$ (top) and Eve's bit error $\varepsilon$ (bottom) with respect to the effective voltage $U_w$ for: $D_2$ (a), $D_3$ (b), and $D_{2,3}$ at $\gamma=10$ (blue), $\gamma=20$ (orange), $\gamma=100$ (yellow), and $\gamma=1000$ (purple). As $\gamma$ and $U_w$ (driven by the effective temperature) decrease, $p$ approaches perfect security.

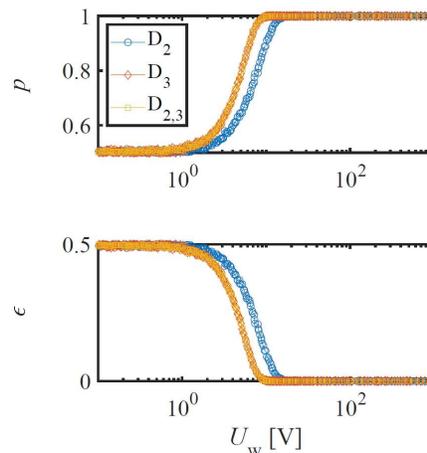

**Figure 5.** Eve's correct-bit-guessing probability $p$ (top) and Eve's bit error $\varepsilon$ (bottom) with respect to the effective voltage $U_w$ at $\gamma = 1,000$ for $D_2$, $D_3$, and $D_{2,3}$. $p$ increases and $\varepsilon$ decreases as $U_w$ (driven by the effective temperature) increases. Convergence to perfect security happens at $D_2$ before $D_3$ and $D_{2,3}$.

## 4. Conclusions

This paper introduces a new passive attack against the KLJN secure key exchange scheme when nonlinearity is present in the transfer function of the amplifier stage of noise generators. We demonstrated the effect of a 1% total distortion at the second order ($D_2$), third order ($D_3$), and a combination of the two orders ($D_{2,3}$) on the KLJN scheme.

We also demonstrated that, at a given nonlinear transfer characteristic, decreasing the effective voltage and, in this way reducing the nonlinearity, is a viable defense against the effect of nonlinearity in the KLJN scheme.

Our results showed that nonlinearity causes a notable power flow that leads to a significant information leak, so a careful design must be implemented such that the total distortion is kept at a minimum.

Alternatively, privacy amplification protocols [43,45,52,93] can also be used. For example, as an active privacy amplification, Alice and Bob can also measure and compare the power flow, and discard a proper fraction of high-risk bits [43,45].